\documentclass[5p,sort&compress]{elsarticle}
\usepackage{graphicx}
\usepackage{txfonts}

\journal{Physics Letters A}

\begin{document}

\begin{frontmatter}

\title{Fluid moment hierarchy equations derived from quantum kinetic theory}

\author{F. Haas, M. Marklund, G. Brodin and J. Zamanian}

\address{Department of Physics, Ume{\aa} University, SE\,--\,901 87 Ume{\aa}, Sweden}

\begin{abstract}
A set of quantum hydrodynamic equations are derived from the moments of the electrostatic mean-field Wigner kinetic equation. No assumptions are made on the particular local equilibrium or on the statistical ensemble wave functions. Quantum diffraction effects appear explicitly only in the transport equation for the heat flux triad, which is the third-order moment of the Wigner pseudo-distribution. The general linear dispersion relation is derived, from which a quantum modified Bohm-Gross relation is recovered in the long wave-length limit. Nonlinear, traveling wave solutions are numerically found in the one-dimensional case. The results shed light on the relation between quantum kinetic theory, the Bohm-de Broglie-Madelung eikonal approach, and quantum fluid transport around given equilibrium distribution functions.
\end{abstract}

\end{frontmatter}

\section{Introduction}

Quantum effects are relevant for macroscopic charged particle systems when the de Broglie wave-length is comparable to the mean inter-particle distance or to the size of the system, when the thermodynamic temperature is comparable to the Fermi temperature or for strong magnetic fields, when spin effects are crucial. In diverse fields such as plasmas \cite{Shukla}, semiconductors \cite{Markowich} or dissipative quantum models \cite{Burghardt}, quantum hydrodynamic equations are important tools due to the relative simplicity in comparison to kinetic theories and the direct physical interpretation of the various quantities involved. Popular ways to include quantum effects in the fluid equations are through the so-called Bohm potential \cite{Gardner, Manfredi} or by means of a quantum spin force \cite{Marklund}, among other generalized forms \cite{Jungel}. In these approaches, some working hypothesis on the underlying quantum statistical ensemble wave functions or the local equilibrium is made. However, there are still issues regarding the relation between the above different approaches that has not been clarified in the literature.

In the present work, we discuss the moment hierarchy equations derived from the electrostatic Wigner equation, which is the quantum counterpart of the Vlasov equation. The moments approach is traditional in classical kinetic theory \cite{Grad}. Without any further constraints or assumptions, a quantum term is explicitly found only at the transport equation for the third-order moment of the Wigner function, with no simplifying assumptions except that we close the system disregarding higher-order moments. The use of moment hierarchy quantum fluid equations is well-known in semiconductor community \cite{Degond, Zhou} but, to the best of our knowledge, was not applied before to the description of electrostatic waves in quantum plasma. The reason for this is that in semiconductor devices there is the presence of a doping profile (an inhomogeneous ionic background) as well as an external heterojunction potential, which makes the analysis intrinsically nonlinear {\it ab initio}. Here we assume a fixed homogeneous ionic background and a mean-field electrostatic potential, so as to obtain a generalized linear dispersion relation, Eq. (\ref{e14}) below. A quantum modified Bohm-Gross dispersion relation is recovered in the long wave-length limit. The nonlinear regimes are numerically analyzed in the one-dimensional case, where traveling-wave solutions are accessible. 

Furthermore, our study is also motivated as a remark on some recent publications \cite{Mostacci}, in which the path from kinetic to fluid quantum models have been reversed. They start from the quantum fluid equations with a Bohm potential term, as derived from the Madelung decomposition of the one-particle wave function, and then insert this back as the momentum change into the Vlasov equation. However, these equations of motion contain the distribution function itself through the averaging procedure, and it is doubtful if they can be used as a basis for kinetic theory.
It can be verified that the resulting dispersion relation disagrees with the Bohm-Pines dispersion relation \cite{Bohm} from quantum kinetic theory, except for zero-temperature equilibria. Since kinetic theories are more accurate than fluid theories, these Bohmian-force methodologies are therefore controversial. See also Ref. \cite{Zheng} for related criticism. The subject bares close connection to the present work on the moment quantum hierarchy, but will be more thoroughly discussed in a separate work.

\section{Governing equations}

Given the one-particle Schr\"odinger equation
\begin{equation}
\label{e0}
	i\hbar\frac{\partial\psi}{\partial t} + \frac{\hbar^2}{2\,m}\nabla^2\psi + e\,\phi\,\psi = 0 \,,
\end{equation}
with the electrostatic potential $\phi$, we can define the Wigner quasi-distribution function according to 
\begin{equation}
\!\!\! \!\!\! \!\!\! \!\!\!\!\! \!	
f({\bf v}, {\bf r}) = \left(\frac{m}{2\pi\hbar}\right)^{3}\! \int \! d{\bf s} \exp[im{\bf v}\cdot{\bf s}/\hbar]\,
	\psi^*(\!{\bf r}+{{\bf s}}/{2})\psi(\!{\bf r}-{{\bf s}}/{2}) .
\end{equation}
Then, from the Schr\"odinger equation, we obtain the quantum kinetic equation
\begin{equation}
\label{e1} 
\frac{\partial f}{\partial t} + {\bf v}\cdot\nabla f + \int d{\bf v}'\,K({\bf v}' - {\bf v},{\bf r})\,f({\bf v}'\!,{\bf r}) = 0 \,, 
\end{equation}
for the Wigner function, coupled to Poisson's equation for the mean-field potential, 
\begin{equation}
\label{e2}
\nabla^{2}\phi = \frac{e}{\varepsilon_0}\left(\int d{\bf v}\,f({\bf v},{\bf r}) - n_{0}\right) \,,
\end{equation}
where $n_0$ is a fixed homogeneous ionic background and $K({\bf v}'-{\bf v},{\bf r})$ is defined by 
\begin{eqnarray}
&& \!\!\! \!\!\! \!\!\! \!\!\! \!\!\! \!\!\! 
K({\bf v}'-{\bf v},{\bf r}) = \frac{i\,e}{\hbar} \left(\frac{m}{2\pi\hbar}\right)^{3} \int\! d{\bf s}\,
\exp\left[{i\,m\,({\bf v} - {\bf v}')\cdot{\bf s}/\hbar}\right]   
\nonumber \\ \label{e3}
&&\qquad \times 
\left[\phi({\bf r}+{{\bf s}}/{2})-
\phi({\bf r}-{{\bf s}}/{2})\right] \,.
\end{eqnarray}
For brevity, the time-dependence of the various quantities is omitted. The Wigner equation (\ref{e1}) holds equally well in the case of mixed states. 

It is interesting to note that a free wave function, given by a Gaussian wave packet, experience a dispersive spreading (as expected), while the corresponding Wigner function does not spread for a fixed value of ${\bf v}$ \cite{Carruthers}. This can also be seen directly from the structure of Eqs.\ (\ref{e0}) and (\ref{e1}). For instance, in the one-dimensional free-particle case and for a Gaussian initial state $\psi(t = 0) = (\sqrt{\pi}\,\sigma)^{-1/2}\,\exp[-x^2/(2\sigma^2)]$, in terms of a variance $\sigma$, the Wigner function can be expressed via $\bar{f} = \exp\left[-(\bar{x}-\bar{v}\,\bar{t})^2 - \bar{v}^2\right]$. Here the rescaled variables $\bar{f} = (\pi\hbar/m)\,f$, $\bar{x} = x/\sigma$, $\bar{v} = m\,v\,\sigma/\hbar$, and $\bar{t} = \hbar\,t/(m\,\sigma^2)$ are employed. The corresponding contour-plot graphics are shown in Fig. 1. 
\begin{figure}
\begin{center}
\includegraphics[width=0.5\textwidth]{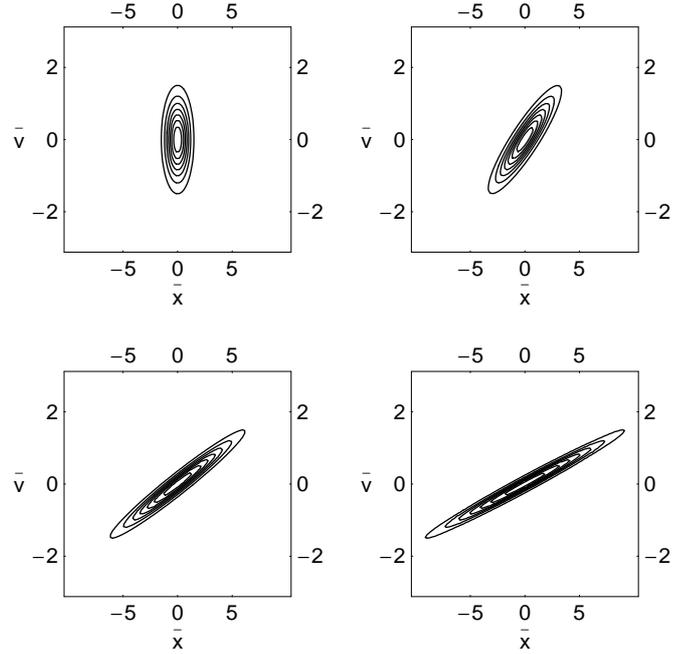}
\caption{Time-evolution of the free-particle rescaled Wigner function $\bar{f} = (\pi\hbar/m)\,f$, in terms of non-dimensional variables $\bar{x} = x/\sigma \,, \bar{v} = m\,v\,\sigma/\hbar$ and $\bar{t} = \hbar\,t/(m\,\sigma^2)$. Initial wave function: $\psi(t = 0) = (\sqrt{\pi}\,\sigma)^{-1/2}\,\exp[-x^2/(2\sigma^2)]$. Upper, left: $\bar{t} = 0$; upper, right: $\bar{t} = 2$; bottom, left: $\bar{t} = 4$;  bottom, right: $\bar{t} = 6$.}
\label{figure1}
\end{center}
\end{figure} 
%

\subsection{Moment equations}

To obtain macroscopic equations, let us introduce the moments
\begin{eqnarray}
\label{e4}
&& \!\!\!\! \!\!\!\! n = \int\! d{\bf v} f \,,\\
\label{e5}
&& \!\!\!\! \!\!\!\! n{\bf u} = \int\! d{\bf v}\,f\, {\bf v} \,,\\
\label{e6} 
&& \!\!\!\! \!\!\!\! P_{ij} = m\,\left(\int\! d{\bf v}\,f\,v_i\, v_j - n\,u_i\,u_j\right) \,, \\
\label{e7}
&& \!\!\!\! \!\!\!\!Q_{ijk} = m\,\int\! d{\bf v}\,(v_i - u_i) (v_j - u_j) (v_k - u_k)\,f \,, \\
\label{e8}
&& \!\!\!\! \!\!\!\!R_{ijkl} = m\,\int\! d{\bf v}\,(v_i - u_i) (v_j - u_j) (v_k - u_k) (v_l - u_l)\,f ,
\end{eqnarray}
from which, in particular, we can derive a scalar pressure $p = (1/3)\,P_{ii}$ and a heat flux vector $q_i = (1/2)\,Q_{jji}$. Here and in the following, the summation convention is applied. The Wigner equation then gives the following macroscopic equations, \\
\begin{eqnarray}
\label{e9}
&& \frac{D\,n}{D\,t} = - n\nabla\cdot{\bf u}  \,, \\
\label{e10}
&& \frac{D\,u_i}{D\,t} = - \, \frac{\partial_j P_{ij}}{m n} + \frac{e}{m}\,\partial_{i}\phi \,, \\
\label{e11}
&& \frac{D P_{ij}}{D\,t} = - \, P_{k(i}\,\partial^{\,k} u_{j)}  - P_{ij}\nabla\cdot{\bf u} - \partial^{\,k} Q_{ij\,k} \,, \\
\label{e12} 
&& \frac{D\,Q_{ijk}}{D\,t} =  \frac{1}{m\,n}\,P_{(ij}\partial^{\,l} P_{k)l}  - \, Q_{l(ij}\,\partial^{\,l} u_{k)}  - Q_{ijk}\nabla\cdot{\bf u} \nonumber \\
&& \qquad\qquad - \frac{e\,\hbar^2 n}{4\,m^2}\,\,\partial^{\,3}_{\,ijk}\phi  - \partial^{\,l} R_{ijkl} \,,
\end{eqnarray}
where $\partial_{\,i} = \delta_{ij}\partial^j = \partial/\partial\,{r_i}$ and the material derivative is $D/D\,t = \partial/\partial\,t + {\bf u}\cdot\nabla$. The calculation assumes {\it e.g.} decaying or periodic boundary conditions and uses the symmetry properties of $P_{ij}$, $Q_{ij\,k}$ and $R_{ijkl}$ under permutation of indices. Finally, in Eqs. (\ref{e11})--(\ref{e12}), the round brackets denote symmetrization, where we use a
minimal sum over permutations of free indicices needed to get symmetric
tensors. Thus, for example, with this convention $P_{k(i}\,\partial^{\,k}u_{j)}$ is defined as $P_{k(i}\,\partial^{\,k}u_{j)}=P_{ki}\,\partial^{\,k}u_{j}+P_{jk}\,\partial ^{\,j}u_{i}+P_{ij}\,\partial ^{\,i}u_{k}$.  

We here remark that a) there is no assumption on the particular equilibrium Wigner function. This is a difference in comparison with some previous approaches \cite{Gardner} relying on the first-order quantum correction to Maxwell-Boltzmann equilibria \cite{Wigner}. Therefore, the model is not semi-classical and is not restricted to classical statistics; b) the explicit dependence on Planck's constant appear only when the heat flux triad transport equation is considered. In addition, the quantum contribution disappears in Eq. (\ref{e12}) when the scalar potential is absent. This is similar to Gardner's approach \cite{Gardner} and in contrast to the method of Ref. \cite{Manfredi}, where quantum effects modeled by a Bohm potential appear already at the momentum transport equation, via the pressure dyad $P_{ij}$ and the associated Madelung decomposition of the quantum statistical ensemble wave functions. Here the usual quantum force is replaced by the third-order derivative of the scalar potential term in Eq. (\ref{e12}). 

It is to be expected from the very beginning that Planck's constant would not appear through the moments of the Wigner equation when there is no electric field, because in this case the Wigner equation reduces to the free-particle Vlasov equation (see the discussion above and Fig.\ 1) and the initial conditions on the Wigner function determines the quantum aspects of its evolution. To conclude, in the present context there are two sources of quantum terms, one via the explicit dependence of the Wigner equation in $\hbar$ and the other via the definition of the Wigner function associated with the proper quantum statistical ensemble.  

\subsection{Closure and dispersion relation}

In this work we choose the simplest way to achieve closure of the system (\ref{e9})--(\ref{e12}), neglecting the contribution from the fourth-order moment $R_{ijkl}$. We take into account Poisson's equation and linearize around the homogeneous equilibrium $n = n_0$, ${\bf u} = 0$, 
\begin{equation}
	[P_{ij}] = n_{0} \kappa_{B} \Bigl[T_{0\perp} ({\bf\hat{x}}\otimes{\bf\hat{x}} + {\bf\hat{y}}\otimes{\bf\hat{y}}) + T_{0\parallel}{\bf\hat{z}}\otimes{\bf\hat{z}}\Bigr] ,
\end{equation}  
$Q_{ijk} = 0$, and $\phi = 0$, where the equilibrium temperatures perpendicular and parallel to the wave propagation  $T_{0\perp}$ and $T_{0\parallel}$ can be unequal. Here, $\kappa_B$ is Boltzmann's constant and plane wave perturbations proportional to $\exp(ik\,z - i\omega t)$ are assumed, without loss of generality. 
It follows that
\begin{equation}
\label{e14}
\omega^2 = \frac{\omega_{p}^2}{2}\,\left[1 + \left(1 + \frac{12\,\kappa_{B}\,T_{0\parallel}\,k^2}{m\,\omega_{p}^2} + \frac{\hbar^2\,k^4}{m^2\,\omega_{p}^2}\right)^{\!1/2}\,\right] \,,
\end{equation}
where $\omega_p = (e^2n_0/m\epsilon_0)^{1/2}$ is the plasma frequency and the equilibrium temperature $T_{0\perp}$ perpendicular to the wave propagation does not contribute.
In the particular case of small wave-vector and quantum effects, Eq. (\ref{e14}) reduces to the usual quantum Langmuir dispersion relation $\omega^2 = \omega_{p}^2 + 3\,\kappa_{B}\,T_{0}\,k^2/m + \hbar^2\,k^4/4\,m^2$. However, even in the formal classical limit ($H \equiv 0$), Eq. (\ref{e14}) yields the Bohm-Gross dispersion relation only for long wave-lengths. Finally, despite the appearance, the new dispersion relation is not restricted to Maxwell-Boltzmann equilibria. For instance, in the case of a zero-temperature, degenerate electron gas, the only basic change would be the substitution of the parameter $T_{0\parallel}$ by the Fermi temperature.  

Contrarily to the habitual usage, a scalar pressure dyad $P_{ij} = p\,\delta_{ij}$ is not a valid assumption even in our electrostatic case. Indeed, linearization of Eqs.~(\ref{e10})--(\ref{e12}) with $P_{ij} = P_{0ij} + \epsilon \delta P_{ij}$, $\phi = \epsilon\delta\phi$, $\epsilon \ll 1$, gives 
\begin{equation}
\label{x}
\delta P_{ij} = - \frac{e\,\delta\!\phi\,k^2}{m\,\omega^2}\,\left(P_{0ij} + P_{0(iz}\delta_{jz)} + \frac{n_{0}\,\hbar^2\, k^2\, \delta_{iz}\,\delta_{jz}}{4\,m}\right) 
\end{equation}
as the first-order perturbation of the pressure dyad, assuming $k_i = k\,\delta_{iz}$ without loss of generality. Clearly the wave propagation itself is a source of anisotropy, even for isotropic equilibria and/or purely classical plasma. The result (\ref{x}) also follows from the kinetic theory in the long wave-length limit. Furthermore, as apparent from Eq.\ (\ref{x}), it is legitimate to postulate 
\begin{equation}
	[P_{ij}] = n \kappa_{B} \Bigl[T_{\perp} ({\bf\hat{x}}\otimes{\bf\hat{x}} + {\bf\hat{y}}\otimes{\bf\hat{y}}) + T_{\parallel}{\bf\hat{z}}\otimes{\bf\hat{z}}\Bigr] , 
\end{equation}
where the perpendicular and parallel temperatures $T_{\perp}$ and $T_{\parallel}$ in general are different. 

We note that the Bohm-Gross dispersion relation is recovered from Eqs.\ (\ref{e9})--(\ref{e11}) in the adiabatic and classic case. Indeed, setting $Q_{ijk} = 0$ and linearizing, the result is $\omega^2 = \omega_{p}^2 + 3\,(\kappa_{B}\,T_{0\parallel}/m)\,k^2$, where only the component $P_{0zz} \equiv n_{0}\,\kappa_{B}\,T_{0\parallel}$ of the equilibrium pressure dyad contributes. On the other hand, insisting on an isotropic pressure dyad $P_{ij} = p\,\delta_{ij}$ and taking $Q_{ijk} = 0$, and combining the continuity equation with Eq. (\ref{e11}) we get $p = n_0 \kappa_B T_0 \,(n/n_{0})^{\gamma}$, where $\gamma = 5/3$, implying the dispersion relation $\omega^2 = \omega_{p}^2 + \gamma\,(\kappa_B T_{0}/m)\, k^2$. To properly recover the Bohm-Gross dispersion relation in an adiabatic, scalar pressure fluid theory retaining only up to the first order moment, there is the need of a phenomenological adiabatic exponent $\gamma = 3$, reflecting the fact that plane wave propagation is essentially a one-dimensional phenomena \cite{Manfredi2}.  Alternative moment hierarchy formulations \cite{Zhou}, closed at the temperature (basically the trace of the second-order moment of the Wigner function) evolution equation, can be shown to result in $\omega^2 = \omega_{p}^2 + (5/3)\,(\kappa_B T_{0}/m)\, k^2 + \hbar^{2}\,k^4/(12\,m^2)$, which goes neither to the Bohm-Gross nor Bohm-Pines dispersion relations in the classical or zero-temperature limits,  respectively. In contrast, as shown here, the quantum modified Bohm-Gross dispersion relation is a natural consequence from third-order moment theory, in the long wave-length limit of Eq. (\ref{e14}).

\subsection{Dynamics}

As an example of the dynamics of the third order hierarchy model, we look at the strictly one-dimensional case, such that ($\partial_{i} = \delta_{ix}\,\partial_{x}$, $u_i = u\,\delta_{ix}$, $P_{ij} = p\,\delta_{ix}\delta_{jx}$, $Q_{ijk} = Q\,\delta_{ix}\delta_{jx}\delta_{kx}$), 
\begin{eqnarray}
\label{e15}
&& \dot{n} + n\,\partial_{x}u = 0 \,, \\
\label{e16}
&& \dot{u} = - \frac{\partial_{x}p}{m\,n} + \frac{e\,\partial_{x}\phi}{m} \,, \\
\label{e17}
&& \dot{p} = - 3\,p\,\partial_{x}u - \partial_{x}Q \,, \\
\label{e18}
&& \dot{Q} = \frac{3\,p\,\partial_{x}p}{m\,n} - \frac{e\,\hbar^2\,n\,\partial^{3}_{x}\phi}{4\,m^2} - 4\,Q\,\partial_{x}u \,,
\end{eqnarray}
and
\begin{equation}
\label{e19}
\partial_{x}^{2}\phi = \frac{e}{\varepsilon_0} (n - n_{0}) \,,
\end{equation}
where the dot represents the convective derivative and disregarding the fourth-order moment. The dispersion relation in Eq. (\ref{e14}) follows from Eqs. (\ref{e15}--\ref{e19}).  

Assuming traveling solutions with all quantities depending only on the variable $\xi = x - v\,t$, where $v$ is a fixed parameter, the following nonlinear system of ordinary differential equations is obtained, 
\begin{eqnarray}
\label{e20}
(u - v)\,u' &=& - \frac{p'}{m\,n} + \frac{e\,\phi'}{m} \,, \\
\label{e21}
(u - v)\,p' &=& - 3\,p\,u' - Q' \,, \\
\label{e22}
(u - v)\,Q' &=& \frac{3\,p\,p'}{m\,n} - \frac{e\,\hbar^2\,n\,\phi^{'''}}{4\,m^2} - 4\,Q\,u' \,, \\
\label{e23}
\phi^{''} &=& \frac{e}{\varepsilon_0}(n - n_{0}) \,,
\end{eqnarray}
where the prime denotes derivative with respect to $\xi$. Finally, the continuity equation can be integrated to $n_0 u_0 = n (u - v) \equiv {\rm cte.}$, in terms of a reference velocity $u_0$ which we assume to be nonzero to exclude trivial cases. 

Eliminating $n$ through the continuity equation, it can be shown that the system of Eqs. (\ref{e20})--(\ref{e23}) admit linearly stable oscillations around the equilibrium $u = u_0 + v, p = p_0, Q = 0, \phi = \phi'=0$, provided the inequality $\hbar\,\omega_{p}/(m\,u_{0}^2) < 2$ is satisfied. Figures 2 and 3 show typical oscillations in this case.  
\begin{figure}[ht]
\begin{center}
\includegraphics[width=0.8\columnwidth]{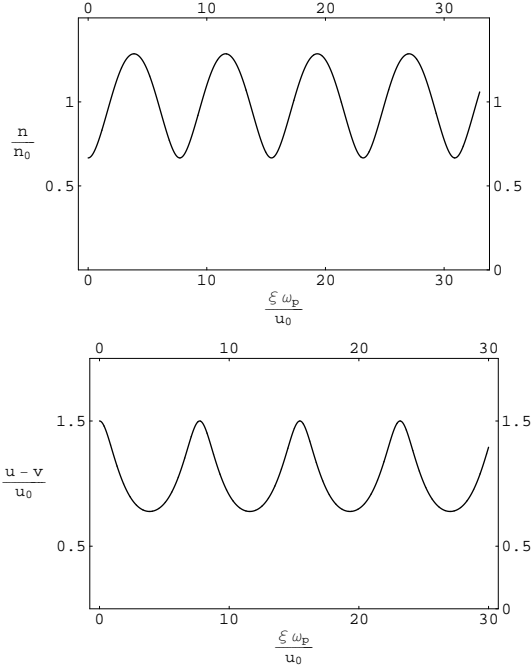}
\caption{Particle density and velocity field oscillations from the system of Eqs. (\ref{e20}--\ref{e23}), where $\xi = x - v\,t$, $u_0 \neq 0$ and for $\hbar\,\omega_{p}/(m\,u_{0}^2) = 1$. Initial conditions such that $n(0) = (2/3)\,n_0, u(0)-v = (3/2)\,u_0, p(0) = m\,n_{0}\,u_{0}^2, Q(0) = 0, \phi(0) = 0, \phi'(0) = 0$.}
\label{figure2}
\end{center}
\end{figure} 
\begin{figure}[ht]
\begin{center}
\includegraphics[width=0.8\columnwidth]{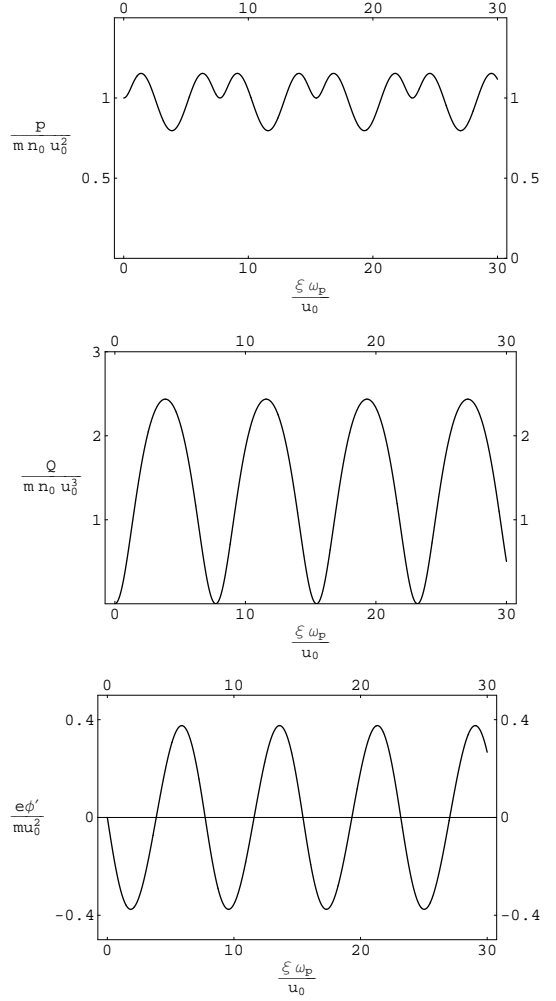}
\caption{Pressure, heat flux and electric field oscillations from the system of Eqs. (\ref{e20}--\ref{e23}), where $\xi = x - v\,t$, $u_0 \neq 0$, and for $\hbar\,\omega_{p}/(m\,u_{0}^2) = 1$. Initial conditions such that $n(0) = (2/3)\,n_0, u(0)-v = (3/2)\,u_0, p(0) = m\,n_{0}\,u_{0}^2, Q(0) = 0, \phi(0) = 0, \phi'(0) = 0$. }
\label{figure3}
\end{center}
\end{figure} 
%

\section{Conclusions}

To conclude, a higher order moment quantum hydrodynamic model for quantum plasmas was derived, starting from the electrostatic Wigner equation. Quantum effects appear explicitly going up to the third-order moments hierarchy, without the need of a Madelung decomposition of the underlying quantum statistical ensemble wave functions or assumptions on the local equilibrium configuration. A generalized dispersion relation for linear waves is derived, from which the quantum modified Bohm-Gross dispersion relation is recovered in the long wave-length limit. For closure, fourth-order moments were discarded. More sophisticated closure schemes \cite{Hammett} designed to reproduce some of the results from kinetic theory, as well as the inclusion of spin effects, are postponed to future considerations. 

\section*{Acknowledgments}
F.H. acknowledges the support provided by Ume{\aa} University and the Kempe Foundations. 
This work is supported by the European Research Council under Contract No.\ 204059-QPQV, and the Swedish Research Council under Contract No.\ 2007-4422.

\end{document}